\documentclass{PoS}

\usepackage{cite}
\usepackage{graphicx}
\usepackage{amssymb,amsfonts,amsmath,bbold,latexsym}

\def\JET{J}
\def\d{\hbox{d}}

\newcommand{\scs}{\scriptscriptstyle}

\def\PR1{
   \parbox[h]{0.15\textwidth}{\includegraphics[width=0.15\textwidth]{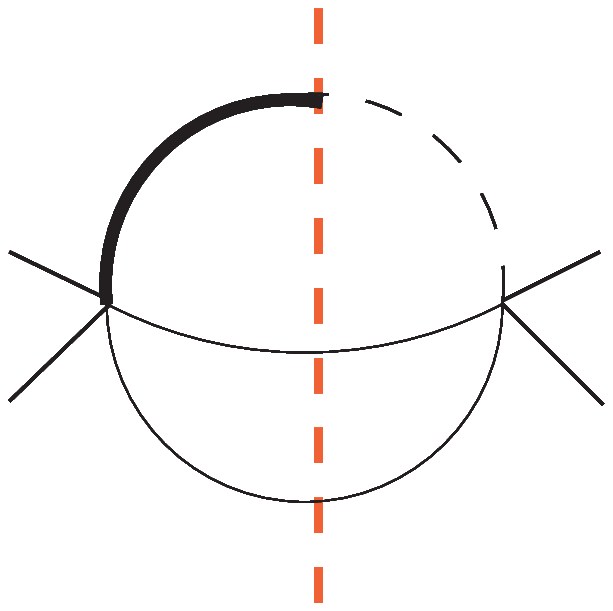}}
}

\title{NNLO antenna subtraction with two hadronic initial states}

\ShortTitle{Initial-initial antennae}

\author{\speaker{Radja Boughezal}
\\
        Institut f\"ur Theoretische Physik,
        Universit\"at Z\"urich,\\
        Winterthurerstr. 190,
        8057 Z\"urich, Switzerland\\
        E-mail: \email{radja@physik.uzh.ch}}
\author{Aude Gehrmann-De Ridder, Mathias Ritzmann\\
Institut f\"ur Theoretische Physik, ETH,\\ CH-8093 Z\"urich,
Switzerland}

\abstract{We discuss the extension of the antenna subtraction method 
to include two hadrons in the initial state 
(initial-initial antennae) at next-to-next-to-leading order\,.
We sketch the construction of the subtraction terms and the required
phase space transformations\,. 
We discuss the integration of the subtraction terms in detail\,. }

\FullConference{RADCOR 2009 - 9th International Symposium on Radiative
Corrections (Applications of Quantum Field Theory to Phenomenology) ,\\
                 October 25 - 30 2009\\
                 Ascona, Switzerland}

\begin{document}
\section{Introduction}
\label{sec:introduction}
Jet observables are an important tool for precision studies due to their large
production cross sections at high energy colliders. 
Calculating perturbative higher order corrections to jet production cross
sections requires a systematic procedure to extract infrared singularities from
real radiation contributions. The latter arise if one or more final state
particles become soft or collinear. At the next-to-leading order level (NLO),
several systematic and process-independent procedures are available\,.
The two main methods are phase space slicing~\cite{Giele:1991vf,Giele:1993dj}
and subtraction based methods~\cite{Frixione:1995ms,Catani:1996vz}.
All subtraction methods consist
in introducing terms that are subtracted from the real radiation part at each   
phase space point\,. These subtraction terms approximate the matrix element in
all singular limits, and are sufficiently simple to be integrated over 
the corresponding phase space analytically\,. After this integration, the
infrared divergences of the subtraction terms become explicit and 
the integrated subtraction terms can be added to the virtual corrections
yielding an infrared finite result\,.  
Since experimental data of jet observables are reaching an accuracy of 
a few percent or better, accurate precision studies must rely on
theoretical predictions that have the same precision\,. 
In some cases, this requires corrections at the next-to-next-to-leading order 
(NNLO) level in perturbative QCD\,. \\

NNLO calculations of observables with $n$ jets in the final state require
several ingredients: the two-loop correction to $n$-parton matrix elements, the
one-loop correction to $(n+1)$-parton matrix elements, and the tree-level
$(n+2)$-parton matrix elements. For most massless jet observables of
phenomenological interest, the two-loop matrix elements have been
computed some time ago, while the other two types
of matrix elements are usually known from calculations of NLO
corrections to $(n+1)$ jet production~\cite{mcfm}\,.
At the NNLO level, $n$ jet observables receive contributions from the one-loop
($n+1)$-parton matrix elements, where one of the involved partons can become
unresolved (soft or collinear), as well as from $(n+2)$-parton matrix elements
where up to two partons can become simultaneously soft and/or collinear\,. 
In order to determine the contribution to NNLO jet observables from these
configurations, two-parton subtraction terms have to be constructed\,. 
Several NNLO subtraction methods have been proposed in the
literature~\cite{Kosower:2002su,nnlosub2,nnlosub3,nnlosub4,nnlosub5}\,.
Another approach used for NNLO calculations of exclusive observables
is sector decomposition~\cite{secdec}\,.  
\\

In ref.~\cite{GehrmannDeRidder:2005cm}, an NNLO subtraction method was
developed for observables with partons in the final state only, the antenna
subtraction method. It constructs the subtraction terms from the so-called
antenna functions. The latter describe all unresolved partonic radiation 
between a hard pair of colour-ordered partons, the radiators. The antenna
functions are derived systematically from physical matrix elements and can
be integrated over their factorized phase space. At the NLO level, 
this formalism can handle
massless partons in the initial or final states~\cite{Daleo:2006xa}, 
as well as massive fermions in
the final state~\cite{ritzmann}. For processes with initial-state 
hadrons, NNLO antenna subtraction terms have to be constructed for two different
cases: only one radiator parton is in the initial state (initial-final
antenna) or both radiator partons are in the initial state 
(initial-initial antenna). Recently, in~\cite{Daleo:2009yj,Luisoni10}, 
NNLO initial-final antenna functions were derived and integrated over their
factorized phase space\,. The case with two radiators in the initial 
state is however still outstanding\,. 
\\

In this contribution, we discuss the derivation of NNLO initial-initial
antenna functions. We briefly describe the construction of the subtraction 
terms and the required phase space transformations and discuss how the phase
space integrals for initial-initial antennae can be performed using
multi-loop techniques\,. 
    
\section{Subtraction terms for initial-initial configurations}
\label{sec:definitions}
At NNLO, there are two types of contributions to $m$-jet observables that 
require subtraction: the tree-level $m+2$ parton matrix elements 
(where one or two partons can become unresolved), and
the one-loop $m+1$ parton matrix elements (where one parton can become
unresolved). 
In the tree-level double real radiation case, we can distinguish four 
different types of unresolved configurations depending
on how the unresolved partons are colour connected to the emitting
hard partons (see ref.~\cite{GehrmannDeRidder:2005cm} 
for a detailed description of the four cases).
In this contribution, we focus on the case with two colour-connected 
unresolved partons (colour-connected). 
This is the only case where new ingredients are needed,
namely the four-parton initial-initial antenna functions. 
The unintegrated ones can be
obtained by crossing two partons to the initial state in the corresponding
final-final antenna functions, which can be found 
in~\cite{GehrmannDeRidder:2005cm},
 and have then to be integrated analytically over
the appropriate antenna phase space\,.    
The corresponding NNLO antenna subtraction term, to be convoluted with the
appropriate parton distribution functions for the initial state
partons, for a configuration with the two hard emitters in the initial state
(partons $i$ and $l$ with momenta $p_1$ and
$p_2$) can be written as:
\begin{eqnarray}
\label{eq:sub2b}
{\rm d}\sigma_{NNLO}^{S,\,\scs{colour-connected}}
&=&  {\cal N}\sum_{m+2}{\rm d}\Phi_{m+2}(k_{1},\ldots,k_{m+2};p_1,p_2)
\frac{1}{S_{{m+2}}} \nonumber \\
&\times& \Bigg [ \sum_{jk}\;\left( X^0_{il,\,jk}
- X^0_{i,\,jk} \,X^0_{Il,\,K} - X^0_{l,\,kj}\, X^0_{iL,\,J} \right)\nonumber \\
&\times&
|{\cal M}_{m}(K_{1},\ldots,{K}_{L},\ldots,K_{m+2};x_1p_1,x_2p_2)|^2\,
\JET_{m}^{(m)}(K_{1},\ldots,{K}_{L},\ldots,K_{m+2})
\Bigg ].
\end{eqnarray}
The subtraction term in eq.~(\ref{eq:sub2b}) is constructed such that
all unresolved limits of the four-parton antenna function $X^0_{il,\,jk}$
are subtracted, so that the resulting subtraction term is active only
in its double unresolved limits, which explains the presence of the products
of three-parton antennae\,.
The subtraction terms for all the other unresolved configurations 
can be constructed using tree-level
three-parton antenna functions. In eq.~(\ref{eq:sub2b})\,, the tree
antennae $X^0_{il,\,jk}$, $X^0_{i,\,jk}$ and $X^0_{l,\,jk}$ depend on the 
original momenta $p_1,\, p_2,\, k_j,\, k_k$\,, whereas the rest of the antenna 
functions as well as the jet function $\JET$ and the reduced 
matrix elements ${\cal M}_m$ depend
on the redefined momenta from the phase space mapping, 
labeled by $I,\,J \dots$. In addition to that, the reduced matrix elements
depend on the momentum fractions $x_1$ and $x_2$\,, which we define later\,.
The normalization factor $\cal N$ includes all QCD-independent factors
as well as the dependence on the renormalized QCD coupling $\alpha_s$, 
$\sum_{m+2}$ denotes the sum over all configurations with $m+2$ partons,
${\rm d}\Phi_{m+2}$ is the phase space for an $(m+2)$-parton final state
in $d=4-2\varepsilon$, and finally,
$S_{m+2}$ is a symmetry factor for identical partons in the final state\,.      
The antenna functions can be integrated analytically, provided we have 
a suitable factorization of the phase space\,. The factorization is possible
through an appropriate mapping of the original set of momenta\,. These 
mappings interpolate between the different soft and collinear limits
that the subtraction term regulates\,. They must satisfy overall momentum 
conservation and keep the mapped momenta on the mass shell\,.\\

A complete factorisation of the phase space into a convolution of an 
$m$ particle phase space depending on redefined momenta only, with 
the phase space of partons $j,\, k$, can be achieved with a Lorentz boost
that maps the momentum $q \;=\; p_1+p_2-k_j-k_k$\,, with $q^2>0$\,, 
into the momentum $\tilde{q} \,=\, x_1 p_1 + x_2 p_2 $\,, where $x_{1,2}$
are fixed in terms of the invariants as follows~\cite{Daleo:2006xa}:      
\begin{eqnarray}
x_1&=&\left(\frac{s_{12}-s_{j2}-s_{k2}}{s_{12}}
\;    \frac{s_{12}-s_{1j}-s_{1k}-s_{j2}-s_{k2}+s_{jk}}{s_{12}-s_{1j}-s_{1k}}
     \right)^{\frac{1}{2}}\,,\nonumber\\
x_2&=&\left(\frac{s_{12}-s_{1j}-s_{1k}}{s_{12}}
\;    \frac{s_{12}-s_{1j}-s_{1k}-s_{j2}-s_{k2}+s_{jk}}{s_{12}-s_{j2}-s_{k2}}
     \right)^{\frac{1}{2}}\,.
\end{eqnarray}
These last two definitions guarantee the overall momentum conservation in the
mapped momenta and the right soft and collinear behavior\,. 
The two momentum fractions satisfy the following limits in
double unresolved configurations:
\begin{enumerate}
\item $j$ and $k$ soft: $x_1\rightarrow 1$, $x_2\rightarrow 1$,
\item $j$ soft and $k_{k}=z_1p_1$: $x_1\rightarrow 1-z_1$, $x_2\rightarrow 1$,
\item $k_{j}=z_1p_1$ and $k_{k}=z_2p_2$:
  $x_1\rightarrow 1-z_1$, $x_2\rightarrow 1-z_2$,
\item $k_{j}+k_{k}=z_1p_1$: $x_1\rightarrow 1-z_1$,
  $x_2\rightarrow 1$,
\end{enumerate}
and all the limits obtained from the ones above by the exchange of $p_1$ with
$p_2$ and of $k_j$ with $k_k$.
The factorized $(m+2)$-partons phase space into an $m$-partons phase space and
an antenna phase space is given by:
\begin{eqnarray}
\d\Phi_{m+2}(k_1,\dots,k_{m+2};p_1,p_2)&=&
\d\Phi_{m}(K_1,\dots,K_{j-1},K_{j+1},\dots,K_{k-1},K_{k+1},\dots,K_{m+2};x_1p_1,x_2p_2)
\nonumber\\
&&\times \; {\cal J} \;\delta(q^2-x_1\,x_2\,s_{12})\,
\delta(2\,(x_2p_2-x_1p_1). q)\,
\nonumber\\
&&\times \;[\d k_j]\;[\d k_k]\;\d x_1\;\d x_2\,,
\label{PS}
\end{eqnarray}
where $[\d k]\, = \,\d^dk/(2\pi)^{(d-1)}\,\delta^+(k^2)$\,, 
and ${\cal J}$ is the Jacobian factor defined by 
\[
{\cal J}=s_{12}\,\left(x_1(s_{12}-s_{1j}-s_{1k}) + x_2
(s_{12}-s_{2j}-s_{2k})\right)\,.
\] 
The next step is to integrate the antenna 
functions over their factorized phase space\,.

\section{Calculational approach for the double real radiation case $2 \rightarrow 3$}
\label{sec:calc}

All the initial-initial antennae have the scattering kinematics 
$p_1 + p_2 \to k_j+k_k + q$, where $q$ is the momentum of the outgoing particle,
for example the vector boson in a vector boson plus jet process\,.
Double real radiation antenna integrals are derived from squared matrix
elements and can be represented by forward scattering diagrams as in 
the following figure:
\vspace{0.2cm}
%
\begin{center}
   \includegraphics[width=0.50\textwidth]{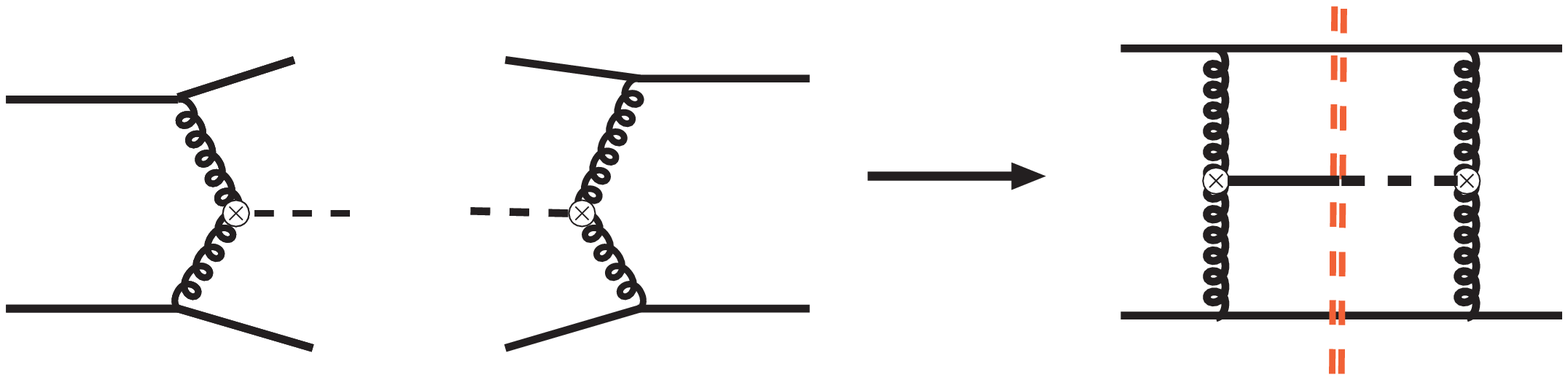}
   \label{fig:cutantennae}
\end{center}
%
The two delta functions in eq.(~\ref{PS}) can be represented  
as mass-shell conditions of fake particles and are shown in the previous
picture as a thick solid line (representing a massive particle with mass 
$M=x_1\,x_2\,s_{12}$) 
and a dashed line (representing a massless particle)\,. 
This allows us to use the optical theorem
to transform the initial-initial antenna phase space integrals into cut
two-loop box integrals and, therefore, use the methods developed for 
multi-loop calculations~\cite{Anastasiou:2002yz,Anastasiou:2003ds}\,. Up
to $8$-propagator integrals with $4$ cut propagators are generated 
in this way\,.
The calculation of the integrated antennae corresponds here 
to the evaluation of a reduced set of master integrals\,. We found 
$30$ of them\,, obtained using integration-by-part (IBP) and 
Lorentz identities, following the Laporta algorithm\,. 
 We then calculate this small set of integrals using the 
method of differential equations\,. The simplest master integral is the two
loop box with all the internal lines cut and defined as follows\\
\vspace{-0.6cm}
\begin{eqnarray}
\label{masterPR1}
&&\PR1 \;\;= \,I(x_1,x_2) \,= \,\int d^d q \,d^dk_j \,d^d k_k\, 
                       \delta^d\left(p_1+p_2-q-k_j-k_k\right) \;
\times \nonumber
\\&& \hspace{4cm}
        \delta^+\left(k_j^2\right) \,
        \delta^+\left(k_k^2\right)\,\delta^+\left(q^2-M^2\right)
        \delta(2\,\left(x_2p_2-x_1p_1).q\right) \,.
\end{eqnarray}
As we have discussed in section~\ref{sec:definitions}, the phase space integrals
(and therefore the master integrals)
have to be studied in four different regions 
of the phase space depending on the values of $x_1$ and $x_2$, namely:
\begin{itemize}
\item $x_1\,\neq\,1$, \,$x_2\,\neq\,1$, we refer to this region as the hard one
\item $x_1\,=\,1$,\, $x_2\,\neq\,1$, and  $x_1\,\neq\,1$,\, $x_2\,=\,1$,
referred to as the collinear region
\item $x_1\,=\,1$,\, $x_2\,=\,1$, is the soft region\,. 
\end{itemize}
In the hard region\,, the solution of the system of differential equations 
yields two-dimensional generalized harmonic polylogarithms\,. 
The $\varepsilon$ expansion is needed up to transcendentality $2$\,. 
In the collinear regions,
additional $1/\varepsilon$ coefficients may be generated and the epsilon 
expansion is done up to transcendentality $3$, whereas in 
the soft region additional 
$1/\varepsilon^2$ coefficients may appear and the expansion in epsilon is 
pushed to transcendentality $4$\,. We note however that the calculation
of the masters in the soft and collinear regions, although needed with deeper 
expansions in $\varepsilon$, is simpler than in the hard
region, and only one-dimensional harmonic polylogarithms are needed in the 
collinear regions\,. In the soft region, a direct calculation is possible 
giving closed form results in $\varepsilon$ in the form of gamma functions\,.  
The boundary conditions for the differential equations are obtained, 
in most of the cases, by studying the master integrals in one of the collinear 
limits\,. Otherwise the soft limit is used\,.
\\

\begin{figure}[h!]
 \begin{center}
   \includegraphics[width=0.70\textwidth]{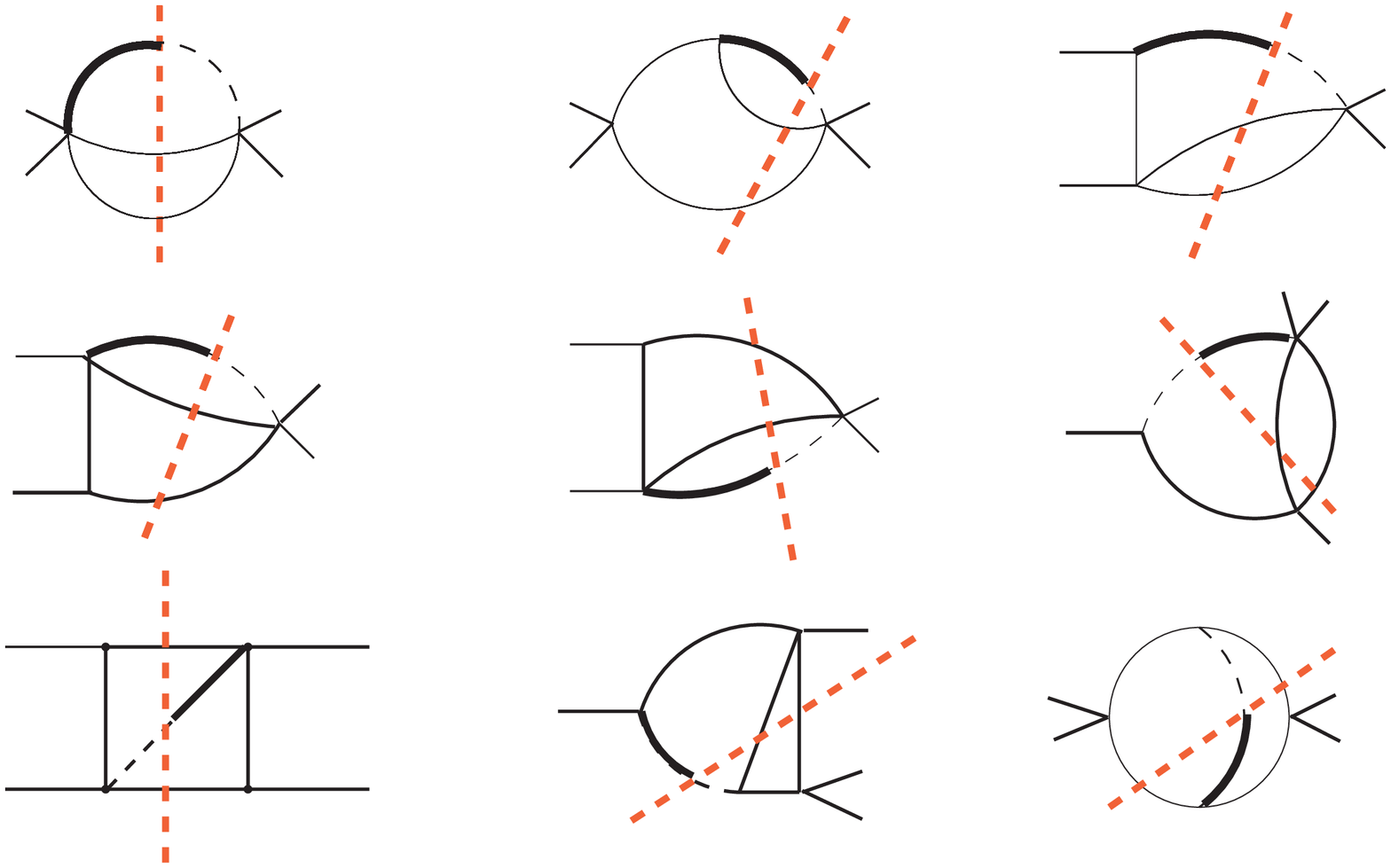}
   \caption{\textsf{Master integrals for the phase space integration 
   of the tree-level initial-initial $B_4^0,\, H_4^0$ and $\tilde{E}_4^0$ 
   type antennae 
   at NNLO\,. Thick solid and
   dashed lines refer to the conditions on the phase space integral
   implemented as auxiliary propagators\,. 
   All the internal lines are massless except for the thick solid line\,.
   Only the integrals without numerators are shown in this picture\,. }}
   \label{fig:mastersHBEt}
 \end{center}
\end{figure}

In a first step towards the calculation of all the integrated 
initial-initial antennae for the $2 \rightarrow 3$ tree-level 
double real radiation case, 
we have focused on all the crossings of two partons from the following
final-final antennae: $B_4^0(q,q',\bar{q}',\bar{q})$, 
$\tilde{E}_4^0(q,q',\bar{q}',g)$ and $H_4^0(q,\bar{q},q',\bar{q}')$ defined
in~\cite{GehrmannDeRidder:2005cm}\,, where the index $4$ refers 
to four partons\,.
There are $13$ master integrals involved in their calculation, 
and the ones
without irreducible scalar products are shown in Fig.~\ref{fig:mastersHBEt}\,.
\\

Finally\,, the one-loop $2 \rightarrow 2$ antenna functions 
do not present any difficulty, since the needed one-loop box
integrals are known analytically to all orders in $\varepsilon$
for the final-final configuration
\,~\cite{Kramer:1986sr}. Obtaining the initial-initial contribution 
using these results requires crossing two legs to the initial state
and performing the necessary analytic continuation of the involved
hypergeometric functions\,. No integrals are required in this case\,.
  
\section{Outlook}
In this contribution, we have discussed the extension of the antenna 
subtraction formalism to the initial-initial configurations\,, including
the required phase space factorisation and mappings\,. We have focused on the 
$2 \rightarrow 3$ tree-level double real radiation contribution\,. 
In a first step
towards the derivation of the complete set of integrated 
initial-initial antennae, we considered all the crossings of 
the subset of $4$-parton antennae:  
$B_4^0(q,q',\bar{q}',\bar{q})$, 
$\tilde{E}_4^0(q,q',\bar{q}',g)$ and $H_4^0(q,\bar{q},q',\bar{q}')$\,\,.
Completing the full set of NNLO antenna functions will allow 
the construction of subtraction terms needed for the evaluation 
of jet observables at hadron colliders\,.
\section{Acknowledgments}
This work is supported by the Swiss National Science Foundation (SNF)
under contracts 200020-116756/2 and PP002-118864\,. 
%


\begin{thebibliography}{99}

\bibitem{Giele:1991vf}
  W.~T.~Giele and E.~W.~N.~Glover,
  Phys.\ Rev.\  D {\bf 46} (1992) 1980.

\bibitem{Giele:1993dj}
  W.~T.~Giele, E.~W.~N.~Glover and D.~A.~Kosower,
  Nucl.\ Phys.\  B {\bf 403} (1993) 633
  [arXiv:hep-ph/9302225].

\bibitem{Frixione:1995ms}
  S.~Frixione, Z.~Kunszt and A.~Signer,
  Nucl.\ Phys.\  B {\bf 467} (1996) 399
  [arXiv:hep-ph/9512328].

\bibitem{Catani:1996vz}
  S.~Catani and M.~H.~Seymour,
  Nucl.\ Phys.\  B {\bf 485} (1997) 291
  [Erratum-ibid.\  B {\bf 510} (1998) 503]
  [arXiv:hep-ph/9605323].



\bibitem{mcfm}
J.~Campbell and R.K.~Ellis,
Phys.\ Rev.\ D {\bf 65} (2002) 113007
[hep-ph/0202176].

\bibitem{Kosower:2002su}
  D.~A.~Kosower,
  Phys.\ Rev.\ D {\bf 67} (2003) 116003
  [hep-ph/0212097].
\bibitem{nnlosub2}
S.~Weinzierl,
JHEP {\bf 0303} (2003) 062
[hep-ph/0302180].
\bibitem{nnlosub3}
W.B.~Kilgore,
Phys.\ Rev.\ D {\bf 70} (2004) 031501
[hep-ph/0403128].
\bibitem{nnlosub4}
M.\ Grazzini and S.\ Frixione,
JHEP {\bf 0506} (2005) 010
[hep-ph/0411399].
\bibitem{nnlosub5}
G.~Somogyi, Z.~Trocsanyi and V.~Del Duca,
JHEP {\bf 0506} (2005) 024
[hep-ph/0502226];

\bibitem{secdec}
C.~Anastasiou, K.~Melnikov and F.~Petriello,
Phys.\ Rev.\ D {\bf 69} (2004) 076010
[hep-ph/0311311];\\
T.~Binoth and G.~Heinrich,
  Nucl.\ Phys.\  B {\bf 585} (2000) 741
  [hep-ph/0004013];\\
G.~Heinrich,
Nucl.\ Phys.\ Proc.\ Suppl.\  {\bf 116} (2003) 368
[hep-ph/0211144];\\
T.~Binoth and G.~Heinrich,
Nucl.\ Phys.\ B {\bf 693} (2004) 134
[hep-ph/0402265].


\bibitem{GehrmannDeRidder:2005cm}
  A.~Gehrmann-De Ridder, T.~Gehrmann and E.~W.~N.~Glover,
  JHEP {\bf 0509} (2005) 056
  [arXiv:hep-ph/0505111].

\bibitem{ritzmann}
 A.~Gehrmann-De Ridder and M.~Ritzmann,
  JHEP {\bf 0907} (2009) 041
  [arXiv:0904.3297].

\bibitem{Daleo:2009yj}
  A.~Daleo, A.~Gehrmann-De Ridder, T.~Gehrmann and G.~Luisoni,
  arXiv:0912.0374 [hep-ph]

\bibitem{Luisoni10}
  A.~Daleo, A.~Gehrmann-De Ridder, T.~Gehrmann and G.~Luisoni,
  arXiv:1001.2397 [hep-ph]\\
  and these proceedings\,.

\bibitem{Daleo:2006xa}
  A.~Daleo, T.~Gehrmann and D.~Maitre,
  JHEP {\bf 0704} (2007) 016
  [arXiv:hep-ph/0612257].

\bibitem{Anastasiou:2002yz}
  C.~Anastasiou and K.~Melnikov,
  Nucl.\ Phys.\  B {\bf 646} (2002) 220
  [arXiv:hep-ph/0207004].

\bibitem{Anastasiou:2003ds}
  C.~Anastasiou, L.~J.~Dixon, K.~Melnikov and F.~Petriello,
  Phys.\ Rev.\  D {\bf 69} (2004) 094008
  [arXiv:hep-ph/0312266].

\bibitem{Kramer:1986sr}
  G.~Kramer and B.~Lampe,
  J.\ Math.\ Phys.\  {\bf 28} (1987) 945.

\end{thebibliography}

\end{document}